# Molecular Nanoelectronics

Dominique Vuillaume

***Abstract*—Molecular electronics is envisioned as a promising candidate for the nanoelectronics of the future. More than a possible answer to ultimate miniaturization problem in nanoelectronics, molecular electronics is foreseen as a possible way to assemble a large numbers of nanoscale objects (molecules, nanoparticules, nanotubes and nanowires) to form new devices and circuit architectures. It is also an interesting approach to significantly reduce the fabrication costs, as well as the energetical costs of computation, compared to usual semiconductor technologies. Moreover, molecular electronics is a field with a large spectrum of investigations: from quantum objects for testing new paradigms, to hybrid molecular-silicon CMOS devices. However, problems remain to be solved (e.g. a better control of the molecule-electrode interfaces, improvements of the reproducibility and reliability, etc…).**

## I. Introduction

Two works paved the foundation of the molecular-scale electronics field. In 1971, Mann and Kuhn were the first to demonstrate tunneling transport through a monolayer of aliphatic chains [1]. In 1974, Aviram and Ratner theoretically proposed the concept of a molecular rectifying diode where an acceptor-bridge-donor (A-b-D) molecule can play the same role as a semiconductor p-n junction [2]. Since that, molecular-scale electronics have attracted a growing interest, both for basic science at the nanoscale and for possible applications in nano-electronics. In the first case, molecules are quantum object by nature and their properties can be tailored by chemistry opening avenues for new experiments. In the second case, molecule-based devices are envisioned to complement silicon devices by providing new functions or already existing functions at a simpler process level and at a lower cost by virtue of their self-organization capabilities, moreover, they are not bound to von Neuman architecture and this may open the way to other architectural paradigms.

Molecular electronics, i.e. the information processing at the molecular-scale, becomes more and more investigated and envisioned as a promising candidate for the nanoelectronics of the future. One definition is "information processing using photo-, electro-, iono-, magneto-, thermo-, mechanico or chemio-active effects at the scale of structurally and functionally organized molecular architectures" (adapted from [3]). In the following, we will review recent results about nano-scale devices based on organic molecules with size ranging from a single molecule to a monolayer. However, problems and limitations remains whose are also discussed.

The structure of the paper is as follows. Section II briefly describes the chemical approaches used to manufacture molecular devices. Section III discusses technological tools used to electrically contact the molecule from the level of a single molecule to a monolayer. Serious challenges for molecular devices remain due to the extreme sensitivity of the device characteristics to parameters such as the molecule/electrode contacts, the strong molecule length attenuation of the electron transport, for instance. Recent advances on these challenges are presented in Section IV. In section V, we discuss the recent progress towards functional molecular devices (e.g. memory, switch,…). It is clear that molecular electronics should be considered as a long term research goal, and that many (if not all) performances of molecular devices cannot compare with more mature CMOS and other less exploratory technologies. Section V gives some highlights on recent results on molecular devices and prototypal chips and discusses some comparisons with other technologies when it makes sense.

## II. Chemistry and Self-Assembly

Making molecular-scale devices requires manipulating and arranging organic molecules on metal electrodes and semiconducting substrates. Organic monolayers and sub-monolayers (down to single molecules) are usually deposited on the electrodes and solid substrates by chemical reactions in solution or in gas phase using molecules of interest bearing a functional moiety at the ends which is chemically reactive to the considered solid surface (for instance, thiol group on metal surfaces such as Au, silane group on oxidized surfaces, etc…) [4]. Many reports in the literature concern self-assembled monoayers (SAMs) of thiol terminated molecules chemisorbed on gold surfaces, and to a less extend, molecular-scale devices based on SAMs chemisorbed on semiconductors, especially silicon. Silicon is the most widely used semiconductor in microelectronics. The capability to modify its surface properties by the chemical grafting of a broad family or organic molecules (e.g. modifying the surface potential [5-7]) is the starting point for making almost any tailored surfaces useful for new and improved silicon-based devices. Between the end of the silicon road-map and the

This work was supported in part by the Centre National de la Recherche Scientifique (CNRS), French Ministry of Research, Agence National de la Recherche (ANR) and the European Union.

D. Vuillaume is with the Molecular Nanostructures and Devices group at the Institute for Electronics Microelectronics and Nanotechnology, CNRS and University of Lille, av. Poincaré, Villeneuve d'Ascq, 59652cedex, France (e-mail: dominique.vuillaume@iemn.univ-lille1.fr).

envisioned advent of fully molecular-scale electronics, there may be a role played by such hybrid-electronic devices [8, 9]. The use of thiol-based SAMs on gold in molecular-scale electronics is supported by a wide range of experimental results on their growth, structural and electrical properties (see a review by F. Schreiber [10]). However, SAMs on silicon and silicon dioxide surfaces were less studied and were more difficult to control. This has resulted in an irreproducible quality of these SAMs with large time-to-time and lab-to-lab variations. This feature may explain the smaller number of attempts to use these SAMs in molecular-scale electronics than for the thiol/gold system. Since the first chemisorption of alkyltrichlorosilane molecules from solution on a solid substrate (mainly oxidized silicon) introduced by Bigelow, Pickett and Zisman [11] and later developed by Maoz and Sagiv [12], further detailed studies [13-16] have lead to a better understanding of the basic chemical and thermodynamical mechanisms of this self-assembly process. For a review on these processes, see Refs. [4, 10]. Moreover, grafting on Si offers more stable devices as compared to thiolated molecules on Au. Due to the labile Au-S bond, SAMs on Au often display current level fluctuations (i.e. random telegraph noise) due to sporadic configurational changes of the Au-S bond [17-19]. This effect is suppressed for SAMs on Si, because the Si-C or Si-O bonds have a larger binding energy [20] and it is only observed for single molecule with more free space around it [21].

Langmuir-Blodgett (LB) monolayers (see a review in a text-book [4]) have also been used in the fabrication of molecular-based devices [22-33], but they are less robust (mechanically and thermally) than SAMs, and chemisorption processes are now more systematically used, both for making monolayers and to attach a single molecule between nano-electrodes. Sublimation of molecules (depending on molecules) can also be used in a sub-monolayer regime especially for UHV-STM studies.

## III. CONTACTING THE MOLECULES

### A. *At the "laboratory" level*

Scanning tunneling microscope (STM) and conducting-atomic force microscope (C-AFM) are widely used at this stage to measure the electronic properties of a very small number of molecules (few tens down to a single molecule). With STM, the electrical "contact" occurs through the air-gap between the molecule or the molecular monolayer and the STM tip (or vacuum in case of an UHV-STM). This leads to a difficult estimate of the true conductance of the molecules [34, 35]. A significant improvement has been demonstrated by Xu and Tao [36] to measure the conductance of a single molecule by repeatedly forming few thousands of Au-molecule-Au junctions. This technique is a STM-based break junction (STM-BJ), in which molecular junctions are repeatedly formed by moving back and forth the STM tip into and out of contact with a gold surface in a solution containing the molecules of interest. A few molecules, bearing two chemical groups at their ends, can bridge the nano-gap formed when moving back the tip from the surface (Fig. 1). Due to the large number of measurements, this technique provides statistical analysis of the conductance data. Using C-AFM as the upper electrode [37-39], the metal-coated tip is gently brought into a mechanical contact with the monolayer surface (this is monitored by the feed-back loop of the AFM apparatus) while an external circuit is used to measure the current-voltage curves. The critical point of C-AFM experiments is certainly the very sensitive control of the tip load to avoid excessive pressure on the molecules [40] (which may modify the molecule conformation and thus its electronic transport properties, or even can pierce the monolayer). On the other hand, the capability to apply a controlled mechanical pressure on a molecule to change its conformation is a powerful tool to study the relationship between conformation and electronic transport [41]. If working on an organic monolayer, an easy technique for a quick assessment of the electrical properties consists in contacting it by a mercury drop [1, 42-45] or a GaIn eutectic drop [46].

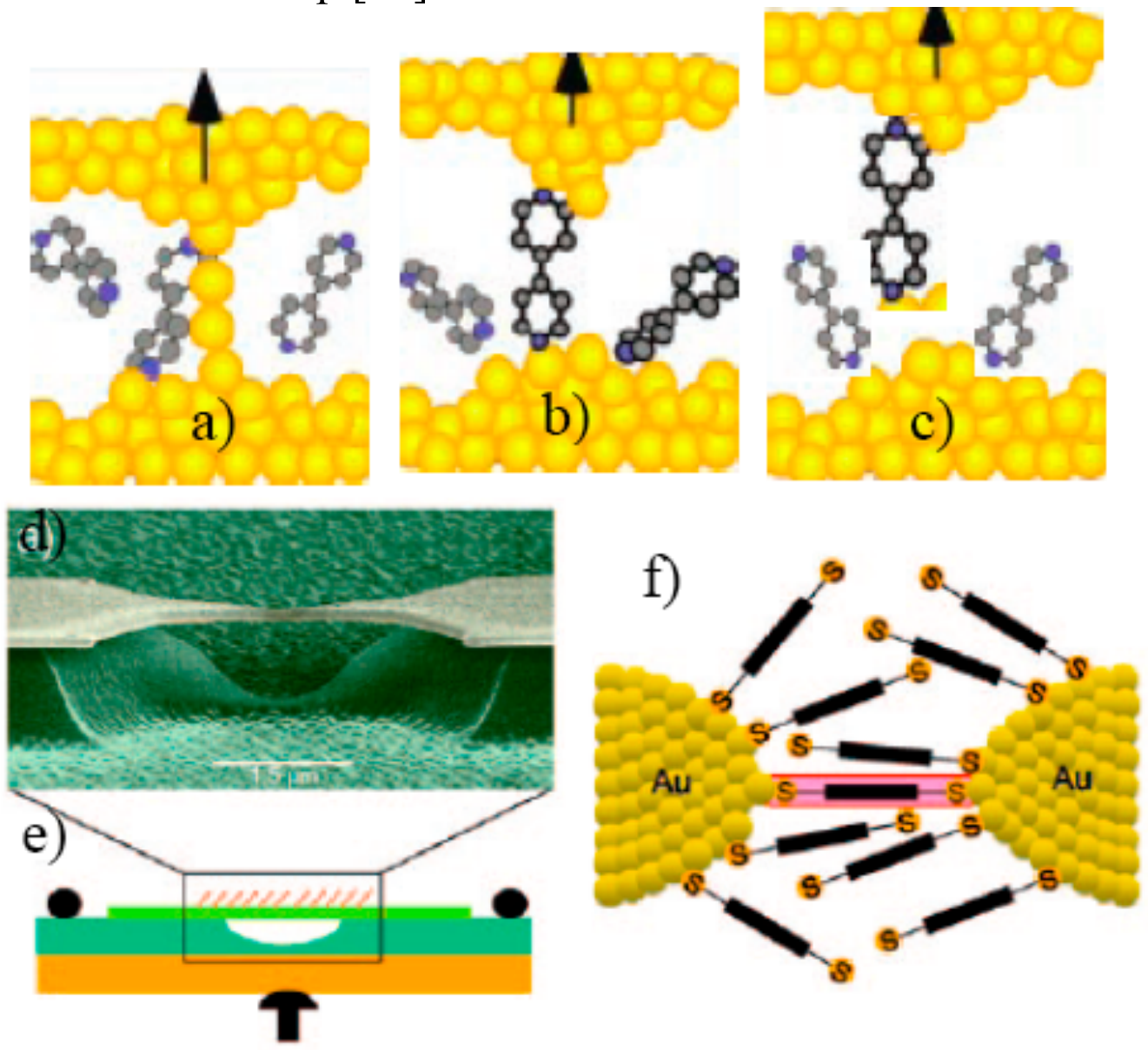

Fig. 1. Typical scheme of the STM break junction with the molecule arrangement while retracting the STM tip [36]. a) Metallic point contact. b) molecule bridging the electrodes. c) no molecule. d) Scanning electron microscope image of a typical mechanically breaking junction [47]. e) Layout of the technique. f) Scheme of the molecule arrangements in the break junction.

Another approach is to use a mechanically breaking junction (MBJ), bridged by few dithiol-terminated molecules [47-51]. A small and suspended metallic nanowire (typically 10 x 10 nm) is fabricated by e-beam lithography on a bendable substrate. A drop of solvent with the molecules of interest is placed on the nanowire, which is elongated and broken by bending the substrate with a piezo system. When the nanowire breaks, a few molecules can chemically bridge the nano-gap and they are simultaneously electrically measured (Fig. 1). Weber et al. reported some improvements allowing stable MBJ measurements at low temperature [52, 53]. Finally, we mention that Au nanoparticles (NP) can be used to connect a few molecules, these NP (tens of nm in diameter) being themselves deposited between electrodes or contacted with a STM [54-56]. A very recent review on how to electrically connect molecules and organic monolayers is given by Haick and Cahen [57]. However, it is clear that connecting molecules with these "laboratory" techniques is not obvious, and in some cases remains at the level of a "tour de force".

### *B. At the "device-like" level*

If we envision device applications, the above techniques are no longer suitable, and the deposition of a metal electrode on top of an organic monolayer, without a degradation of the monolayer and without the creation of metallic shorts, is a critical issue. Several studies [58-63] have analyzed (by X-ray photoelectron spectroscopy, infra-red spectroscopy,…) the interaction (bond insertion, complexation…) between the metal atoms and the molecules. When the metal atoms are strongly reactive with the end-groups of the molecules (e.g. Al with COOH or OH groups, Ti with $COOCH_3$, OH or CN groups….) [58-63], a chemical reaction occurs forming a molecular overlayer on top of the monolayer. This overlayer made of organometallic complexes or metal oxides may perturb the electronic coupling between the metal and the molecule, leading, for instance, to partial or total Fermi-level pinning at the interface [64]. In some cases, if the metal chemically reacts with the end-group of the molecule (e.g. Au on thiol-terminated molecules), this overlayer may further prevent the diffusion of metal atoms into the organic monolayer [65]. The metal/organic interface interactions (e.g. interface dipole, charge transfer,…) are very critical and they have strong impacts on the electrical properties of the molecular devices. Some reviews are given in Refs. [66, 67]. If the metal atoms are not too reactive (e.g. Al with $CH_3$ or $OCH_3$…) [58-63], they can penetrate into the organic monolayer, diffusing to the bottom interface where they can eventually form an adlayer between this electrode and the monolayer (in addition to metallic filamentary short circuits). In a practical way for device application using organic monolayers, the metal evaporation is generally performed onto a cooled substrate (~100 K). It is also possible to intercalate blocking baffles on the direct path between the crucible and the sample, or/and to introduce a small residual pressure of inert gas in the vacuum chamber of the evaporator [27, 28, 68]. These techniques allow reducing the energy of the metal atoms arriving on the monolayer surface, thus reducing the damages.

To avoid these problems, alternative and soft metal deposition techniques were developed. One called nanotransfer printing (nTP), has been described and demonstrated [69]. Nanotransfer printing is based on soft lithographic techniques used to print patterns with nanometric resolution on solid substrates [70]. The principle is briefly described as follows (Fig. 2). Gold electrodes are deposited by evaporation onto an elastomeric stamp and then transferred by mechanical contact onto a thiol-functionalized SAM. Transfer of gold is based on the affinity of this metal for thiol function –SH forming a chemical bond Au–S. Loo et al. [69] have used the nTP technique to deposit gold electrodes on alkane dithiol molecules self-assembled on gold or GaAs substrates. Nanotransfer printing of gold electrodes was also deposited onto oxidized silicon surface covered by a monolayer of thiol-terminated alkylsilane molecule [71, 72]. Soft depositions of pre-formed metal electrodes, e.g. lift-off float-on (LOFO) [73], have also been developed. Recently, a very elegant solution to avoid the formation metallic filamentary paths within the SAM has been proposed in which a thin conducting polymer layer (PEDOT:PSS, poly-ethylene-dioxythiophene) stabilized with poly-styrene-sulphonic acid) has been intercalated as a buffer layer between the organic monolayer and the evaporated metal electrode [74]. With this technique, it is possible to manufacture molecular junctions with a large area (diameter up to 100 μm), a very high yield (> 95%), and with an excellent stability and reproducibility. This simple approach is potentially low-cost and suitable for practical molecular electronics. It was also reported to use metallic electrode made of a 2D network of carbon nanotubes [75].

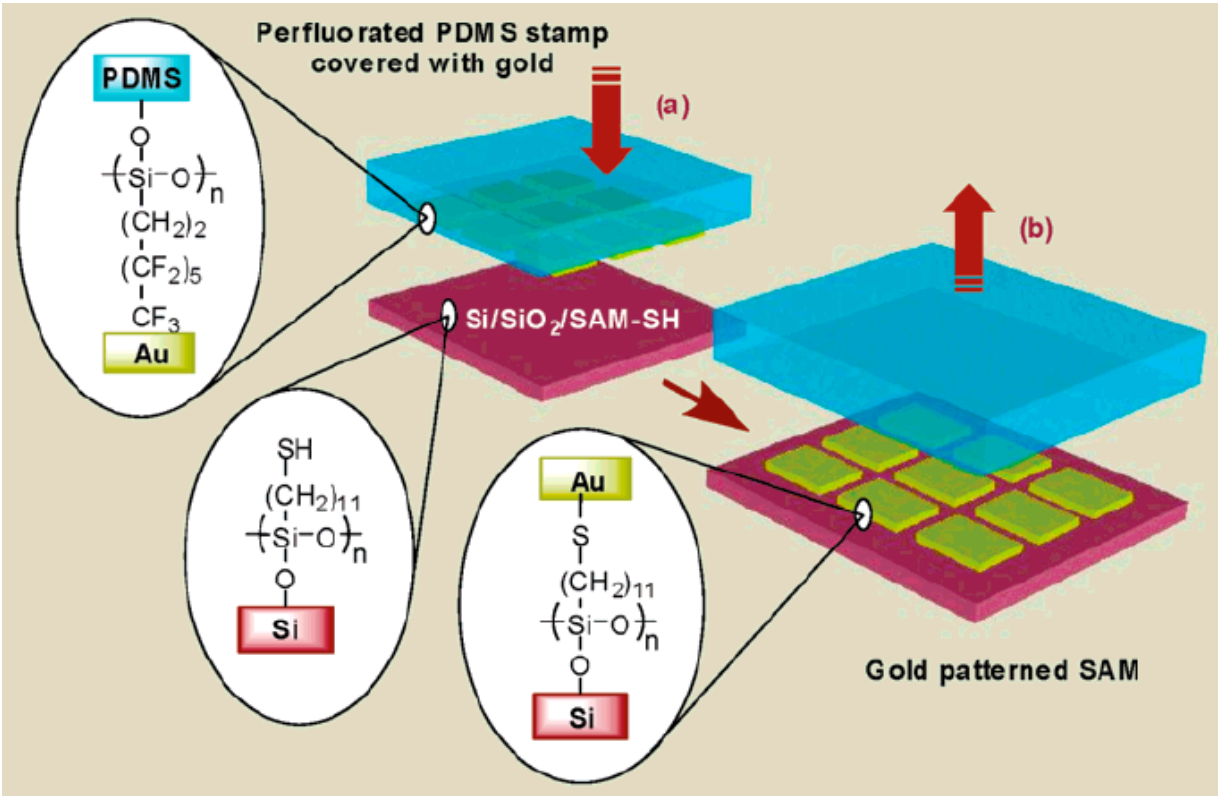

Fig. 2. Principle for deposition of gold electrodes on a thiolated SAM on silicon by nTP method. (a) Bring gold-coated patterned stamp into contact with the SH-functionalized SAM. (b) Remove the stamp from the substrate. Gold electrodes are transferred on the substrate [69, 71].

A transistor structure was also investigated (3-terminal device) using a bottom gate transistor configuration. The difficulties are (i) to make these electrodes with a nanometer-scale separation; (ii) to deposit molecules into these nano-gaps. Alternatively, if the monolayer is deposited first onto a suitable substrate, it would be very hard to pattern, with a nanometer-scale resolution, the electrodes on top of it. The monolayers have to withstand, without damage, a complete electron-beam patterning process for instance. This has been proved possible for SAMs of alkyl chains [76, 77] and alkyl chain functionalized by π-conjugated oligomers [78] used in nano-scale (15 – 100 nm) devices. However, recently developed soft-lithographies (micro-imprint contact…) can be used to pattern organic monolayers or to pattern electrodes on these monolayers [70]. Nowadays, 30 nm width nano-gaps are routinely fabricated by e-beam lithography and 5 nm width nano-gaps are attainable with a lower yield (a few tens %) [79-81]. However, these widths are still too large compared to the typical molecule length of 1-3 nm. The smallest nanogaps ever fabricated have a width of about 1 nm. A metal nanowire is e-beam fabricated and a small gap is created by electromigration when a sufficiently high current density is passing through the nanowire [82]. These gold nanogaps were then filled with few molecules (bearing a thiol group at each ends) and Coulomb blockade and Kondo effects were observed in these molecular devices [83, 84]. A second approach is to start by making two electrodes spaced by about 50-60 nm, then to gradually fill the gap by electrodeposition until a gap of few nanometers has been reached [85-87]. Recently, carbone nanotubes (CNT) have been used as electrodes separated by a nano-gap (<10 nm) [88]. The nano-gap is obtain by a precise oxidation cutting of the CNT, and the two facing CNT ends which are now terminated by carboxylic acids, are covalently bridged by molecules of adapted length derivatized with amine groups at the two ends (Fig. 3). It is also possible to functionalize the molecule backbone for further chemical reactions allowing the electrical detection of molecular and biological reactions at the molecule-scale [88, 89].

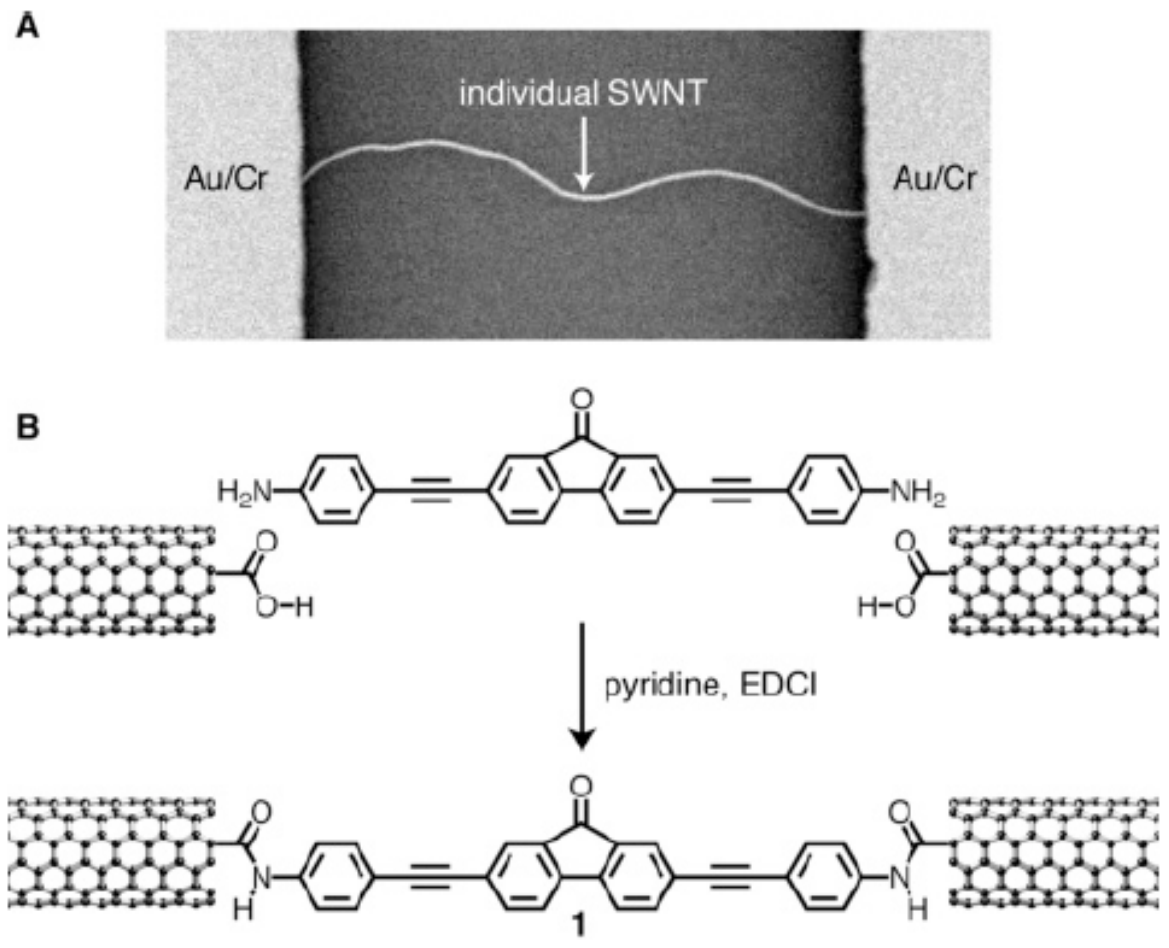

Fig. 3. A) SEM image of a CNT before cutting and bridging with a molecule. B) Scheme of the CNT-molecule-CNT junction [88, 89].

## IV. THE MOLECULE/ELECTRODE CONTACT CHALLENGE

As discussed above, it is clear that the difficulties of making and controlling electronic connections to molecules are the great challenge. It has long been recognized that the electrical conductance of a molecular junction (i.e. a molecule or monolayer sandwiched between two electrodes, whether they are metallic or semiconducting) is strongly influenced by the chemical nature and atomic configuration of the molecule/electrode contact. Small changes in the contact geometry can dramatically change the conductance through the molecule [90]. For instance, theoretical calculations have predicted that selenium (Se) and tellurium (Te) are better links than sulfur (S) for the electronic transport through molecular junctions [91, 92]. This was demonstrated in a series of experiments using SAMs made of bisthiol- and biselenol-terthiophene molecules (a π-conjugated molecule prototype of a "molecular wire") inserted in a dodecanethiol matrix (forming an electrically insulating matrix because alkane molecules have a large HOMO-LUMO gap) [93, 94]. Further experiments have shown that: i) amine group ($NH_2$) give better controlled conductance variability than thiol (SH) and isonitrile (CN) [95] and ii) the interface contact resistance is lower for amine than for thiol [96]. For further reading on the physics of molecule/electrode contacts, on the influence of the contact on molecular energetics and how it impacts electron transport phenomenon in molecular devices, the interested reader can found more in recent review papers [97-99].

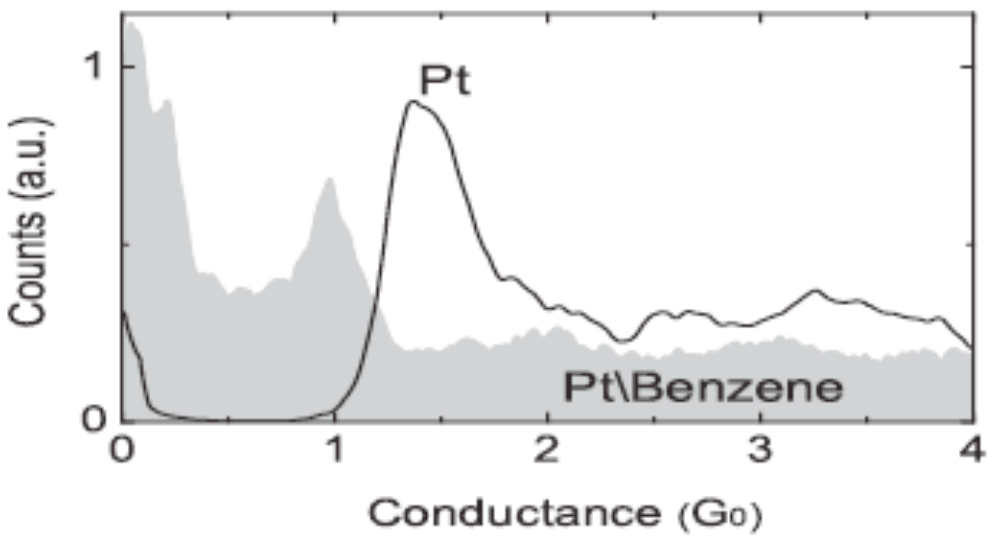

Fig. 4. Histogram of the conductance of a Pt/benzene/Pt junction and comparison with a Pt-Pt point contact [100].

However, this chemical link acts also as a tunnel barrier for electron transfer between the electrode and the molecule [101]. As a consequence, the conductance of a molecular junction is usually small (typically below $10^{-2}$ $G_0$, where $G_0=2e^2/h=77.5\mu S$ is the conductance quantum)[49, 51, 95, 102-104]. Consequently, molecular electronics is a “high impedance” electronics which implies a large power dissipation if we envision a high-density of molecular devices in a same chip (related to the small size of molecules). Recently, a significant progress was made towards “low impedance” molecular electronics. The group of van Ruitenbeek reported on a highly conductive molecular junction, around $G_0$, obtained with a direct binding of small organic molecules (benzene, acetylene, CO, $CO_2$, $H_2$, $H_20$) to metallic electrodes (Pt) without the use of anchoring groups (Fig. 4) [100, 105]. This result has been ascribed to the good reactivity of the Pt, forming direct bonds with the molecule. Although all these molecular junctions have about the same conductance near $G_0$, the contribution of the molecule in the transport properties of the junction has been also evidenced by the observation of different vibrational signatures in the inelastic electron tunneling transport [105].

Further experiments are now required to determine to which extent the conclusions drawn for a peculiar molecule and metal electrode are valid for any other ones. With all these data on hands, one would optimize the design of future devices for molecular electronics.

Another issue is the strong dependence of the conductance of a molecular junction with the length of the molecule. Due to relatively large energy offset at the molecule/electrode interface (see above), the conductance is mainly dominated by tunneling and the conductance follows a classical exp(-βd) law (d is the molecular length) with $\beta \approx 1$ $Å^{-1}$ for saturated molecules (alkyl chain) and ≈ 0.5-0.6 $Å^{-1}$ for π-conjugated molecules.[102] It means that electron transport is limited to few nm in these systems. This limitation has been recently overcome. By introducing, step-by-step, metallic ions (Co(II), Fe(II)) between π-conjugated oligomers (terpyridine-based), the group of M.A. Rampi and coworkers have built "long" molecular wires (up to 40 nm) with a very low attenuation factor (β ≈ 0.001 $Å^{-1}$) (Fig. 5) [106]. This result is due to the introduction of energy levels, related to the metal ions, in close resonance with the Fermi energy of the metal electrodes. While this approach is not completely new (see a review in [107]), this result opens interesting perspectives for the development of molecular wires.

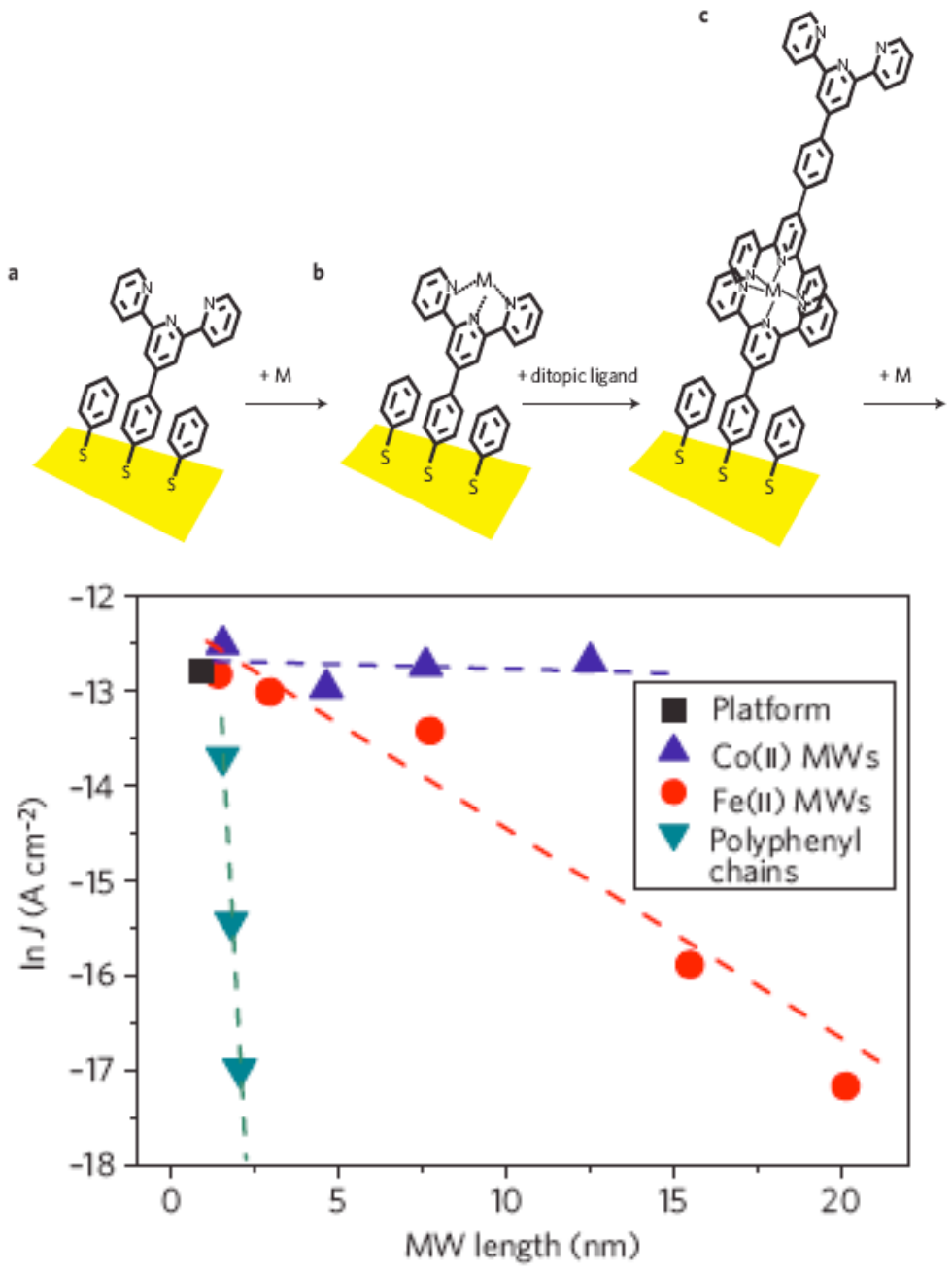

Fig. 5. Schematic representation of the synthesis of the molecular wires, and length dependence of the current for the molecular wires with the Co and Fe ions. Comparison with classical π-conjugated molecules (polyphenyl chains) [106].

Compared to metal nanowires, the resistance of the above mentioned molecular wire remains large (MΩ range), while taken a typical resistivity of 10 μΩ.cm [108], metallic nanowires of the same size have resistances in the range of kΩ. This feature is probably due to a lack of molecule-contact optimization as discussed above. Other groups have developed different chemical approaches (e.g. based on oligoyne derivatives), reporting low attenuation factors (0.06 $Å^{-1}$) [109]. Other theoretical proposals rely on the "doping" of molecules by heteroatoms so as to align the molecular orbitals with the Fermi energy of the metal electrodes [110].

## V. Functional Molecular Devices

### A. *Charge-based memory*

Redox-active molecules, such as mettalocene, porphyrin and triple-decker sandwich coordination compounds attached on a silicon substrate have been found to act as charge storage molecular devices [111-114]. The molecular memory works on the principle of charging and discharging of the molecules into different chemically reduced or oxidized (redox) states. It has been demonstrated that porphyrins (i) offer the possibility of multibit storage at a relatively low potentials (below ~ 1.6 V), (ii) can undergo trillions of write/read/erase cycles, (iii) exhibit charge retention times that are long enough (minutes) compared with those of semiconductor DRAM (tens of ms) and (iv) are extremely stable under harsh conditions (400°C – 30 min) and therefore meet the processing and operating conditions required for use in hybrid molecule/silicon devices [114]. Due to the high density of molecules on the surface (up to $10^{13}$-$10^{14}$ $cm^{-2}$) a high charge density (10-16 μC/$cm^2$) is obtained [114] without the need of complicated device structures (deep-trench and stacked capacitances) as in classical DRAM technologies. The feasibility of a 1Mbit hybrid (molecule on CMOS platform) DRAM has been demonstrated that uses 1/10$^{th}$ of the capacitor area of conventional technology [115]. Moreover, the same principle works with semiconducting nanowires dressed with redox molecules in a transistor configuration [116-118]. Optoelectronic memories have also been demonstrated with polymer-functionalized CNT transistors [119, 120]. However, in all cases, further investigations on the search of other molecules and, understanding the factors that control parameters such as, charge transfer rate, which limit write/read times, and charge retention times, which determines refresh rates, are needed. For instance, the length and the chemical nature of the linker between the redox molecule and the silicon must be adjusted to tune the electrical properties of the device, such as charging and discharging kinetics and retention time [121, 122].

### B. *Configurational switch and memory*

One of the most interesting possibilities for molecular electronics is to take advantage of the soft nature of organic molecules. Upon a given excitation, molecules can undergo configurational changes. If two different configurations are associated with two different conductivity levels of the molecule, this effect can be used to make molecular switches and memories. Such an effect is expected in π-conjugated oligomers if one of the monomer is twisted away from a planar configuration of the molecule. Twisting one monomer breaks the conjugation along the backbone, thus reducing the charge transfer efficiency along the molecule. It has been verified that the conductance follows the expected law $G=G_0\cos^2\theta$, where θ is the torsion angle between the monomers (Fig. 6) [104, 123]. While the molecules are not exactly the same (various lateral substituents are used to impose the tilt through stearic hindrance), the intrinsic role of these substituents on the electron transport properties of the molecule is negligible compared to the configurational change of the bi-phenyl backbone [104, 123].

Catenane and rotaxane are a kind of molecules exhibiting a bistable behavior. In brief, these molecules are made of two parts, one allowed to move around or along the other one (e.g. a ring around a rod, two interlocked rings). These molecules adopt two different configurations depending on their redox states, changing the redox state triggers the displacement of the mobile part of the structure to minimize the total energy. This kind of molecules was tentatively used to build molecular memories. A voltage pulse of about 1.5 - 2 V was used the switch the device from the "off" state to its "on" state. The state was read at a low bias (typically 0.1-0.2 V). The on/off ratio was about a few tens. A pulse in reverse bias (-1.5 to -2 V) returned the device to the "off" state. Using these molecular devices, Chen and coworkers [32, 33] have demonstrated a 64 bits non-volatile molecular memory cross-bar with an integration density of 6.4 Gbit/$cm^2$ (a factor ~10 larger than the state-of-the-art today's silicon memory chip). The fabrication yield of the 64 bits memory is about 85%, the data retention is about 24 h and about 50-100 write/erase cycles are possible before the collapse of the on/off ratio to 1. Recently a 160 kbit based on the same class of molecules has been reported, patterned at a 33 nm pitch ($10^{11}$ bits/$cm^2$) [124]. About 25% of the tested memory points passed an on/off ratio larger than 1.5 with an average retention time of ~ 1h. However, it has also been observed that similar electrical switching behaviors can be obtained without such a class of bistable molecules (i.e. using simple alkyl chains instead of the rotaxanes) [125]. In this latter case, the switching behavior is likely due to the formation and breaking of metallic micro-filaments introduced though the monolayer during the top metal evaporation. The presence of such filaments is not systematic and simple techniques, such as the use of a buffer film of a conducting PEDOT:PSS polymer, have been developed to avoid this “metallic filament” issue (see above, section III), however caution has to be taken before to definitively ascribe the memory effect as entirely due to the presence of the molecules. It is likely that many switching and memory effects reported by many authors are just due to some metallic filaments effect, that would have been avoided using such PEDOT:PSS layer. The advantage of such molecular cross-bar memories are i) a low cost, ii) a very high integration density, iii) a defect-tolerant architecture, iv) an easy post-processing onto a CMOS circuitry and v) a low power consumption. For instance, it has been measured that an energy of ~50 zJ (or ~ 0.3 eV) is sufficient to rotate the dibutyl-phenyl side group of a single porphyrin molecule [126]. This is ~$10^4$ lower than the energy required to switch a state-of-the art MOSFET, and near the kTLn2 (2.8 zJ at 300K, or 0.017 eV) thermodynamic limit.

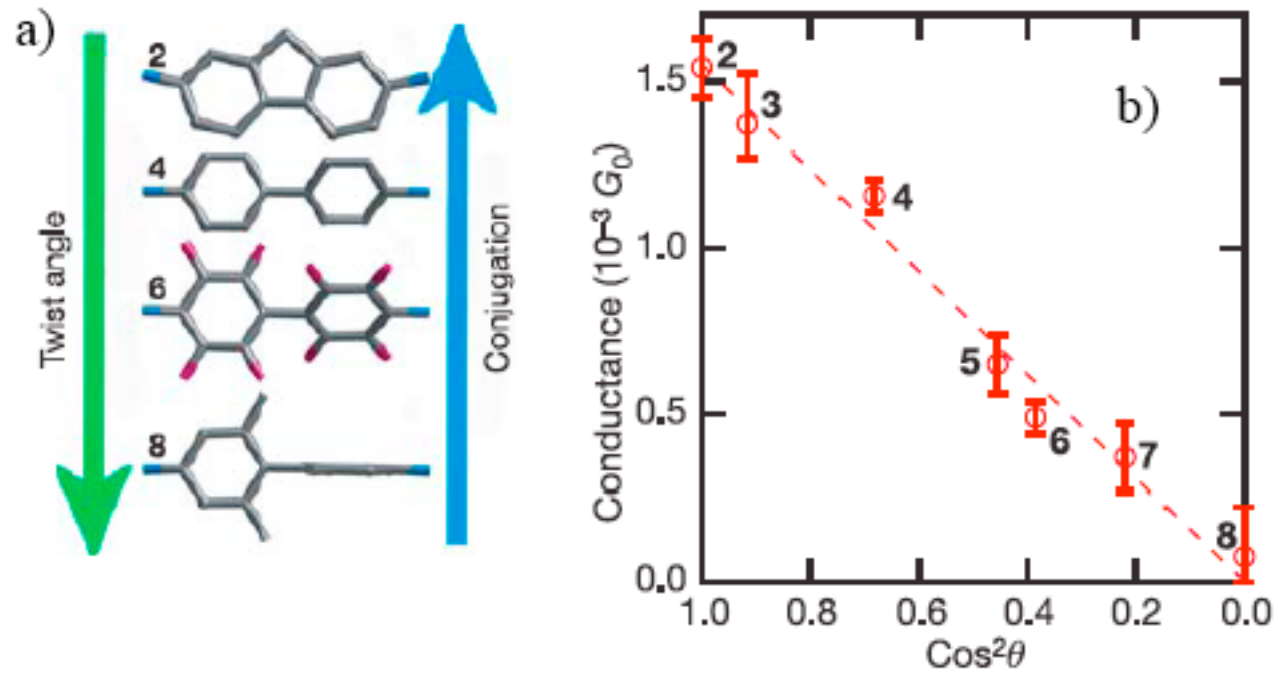

Fig. 6. a) Bi-phenyl based molecules with different torsion angle between the two phenyl rings. b) Measured conductance with the STM break junction versus the torsion angle [104].

Alternative technologies for resistive memories [127] include, for instance, polymer-based memory [128], nanomechanical memory based on NEMS with CNT, graphene [129], nanothermal memory such as nanowire PCM (phase change memory) [130] using phase transformation between amorphous and crystalline phases as in more conventional PCM. All these technologies have demonstrated data retention times of a few months, and write/erase times in the range of ns to ms, the molecular one being the slowest at the moment. This is probably not the end of the story; chemical reactions (reduction-oxidation, configuration switching) of molecules in solution can be fast, and the actual limitations likely reflect the lack of device optimization (e.g. bad control of molecule-electrode contact, etc…) as discussed above.

Another efficient way to trigger a configuration change in a molecule is by light. Reversible photoswitching devices were demonstrated with diarylethene and azobenzene derivatives [131, 132], and open the route for potential applications as optical switches in molecular electronics [133-138]. For instance, azobenzene molecules show a transition from a more thermodynamically stable *trans* configuration to a *cis* configuration upon exposure to UV light (~ 360 nm), and a reversible isomerization under blue light (~ 480 nm). The properties of azobenzene in solution (e.g. robust reversible photoisomerization, long living states, fast switching) make them promising building blocks for molecular-scale devices and nanotechnology. The challenge consists in defining strategies to use these promising materials at the molecular-scale on the surface of electrodes for applications as molecular-switches and memories for instance.

It is well known that photoisomerisable molecules need to be electronically decoupled from the metal surface to properly work, i.e. to reversibly switch between the two isomers. STM experiments on a single azobenzene molecule physisorbed on gold show that reversible switching is only observed when tert-butyl legs lift the molecule up [138]. In several works using SAMs, the photoisomerisable molecules are chemically attached to the substrate using various spacers: short alkyl chains (2 to 6 carbon atoms) [133, 139, 140], ethylene bond [137] or phenyl or thiophene moieties [132, 134, 138, 141, 142]. The role of this linker is crucial. In the case of diarylethene, the observation of reversible switching depends on its nature (e.g. phenyl vs. thiophene) [142]. Contradictory, some authors associate the higher "on" conductance state to the *trans* isomer [133, 137, 140, 143-146], while others conclude in favor of the *cis* form [134, 138, 147]. In all these results, the "on/off" conductance ratio is lower than 50 (Fig. 7-a). Moreover, up to now, azobenzene derivatives do not exhibit a clear intrinsic conductance switching. The apparent change in the measured conductance has been attributed to a change in the length of the molecule during the isomerization rather than to an intrinsic conductance switching associated with changes in the electronic structure of the molecular junction [134, 138]. Many reasons can explain these results; details on the molecular arrangements in the monolayers and the nature of the coupling between the molecule and the electrode contact are among the most important factors that can influence the electrical behavior. For instance, Cuniberti et al. [143, 148] showed by first principles calculations that $G_{trans} > G_{cis}$ when the azobenzene is chemically linked between two carbon nanotube electrodes, while $G_{cis} > G_{trans}$ in case of silicon electrodes ($G_{trans}$ and $G_{cis}$ are the conductances of the junction for the *trans* and *cis* isomers, respectively). Obviously, the role of the spacer is also critical. A short spacer should favor the electron transfer rate through the junction and increases its conductance, while a longer spacer could improve the decoupling of the azobenzene moiety from the substrate, thus allowing a larger dynamic of the switching event, and thus a larger "on/off" conductance ratio.

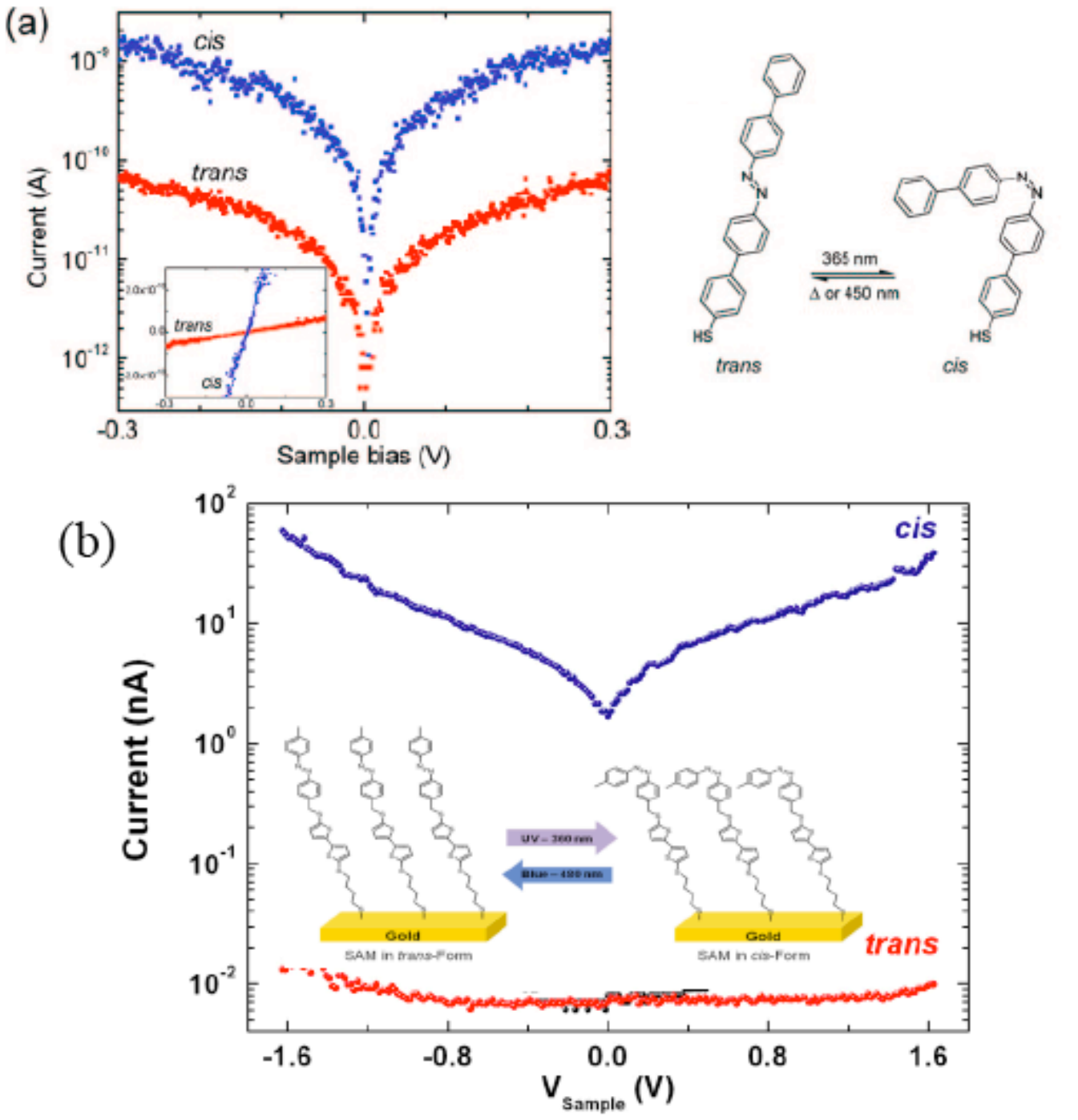

Fig. 7. Current-voltage curves (C-AFM measurements) of the SAM of two types of azobenzene derivatives in the trans and cis conformations: a) Ref. [134] b) Refs. [149, 150].

Recently, we reported the synthesis and the electrical properties of a new molecular switch in which the azobenzene moiety is linked to a bithiophene spacer and a short (4 carbon atoms) alkanethiol [149, 150]. Such a design is expected to combine the benefit of a rather long spacer, while preserving a sufficiently high level of current due to the presence of electron-rich bithiophene unit (compared to a fully saturated spacer with the same length). A record on/off ratio up to $7 \times 10^3$ between the *cis* ("on") and *trans* ("off") configurations was demonstrated (Fig. 7-b). The analysis of these results using well-established electron transport models and molecular frontier orbitals from first principles DFT calculations indicates that this high photo-induced on/off ratio results from a synergistic combination of SAM thickness variation and modification of the energy offset between the lowest unoccupied molecular orbital (LUMO) and the electrode Fermi energy. Moreover, these azobenzene derivatives can switch their configuration with the top electrode deposited on the SAM [138] and they can have switching times ~ 1-10 µs comparable to molécules in solution [150], thus they are prone for solid-state molecular switch devices for low-demanding, low-cost applications.

### *C. Molecular transistors*

A true transistor effect (i.e. the current through 2 terminals of the device controlled by the signal applied on a third terminal) embedded in a single three-terminal molecule (e.g. a star-shaped molecule) has not been yet demonstrated. Up to date, only hybrid-transistor devices have been studied. The typical configuration consists of a single molecule or an ensemble of molecules (monolayer) connected between two source and drain electrodes separated by a nanometer-scale gap, separated from an underneath gate electrode by a thin dielectric film (Fig. 8).

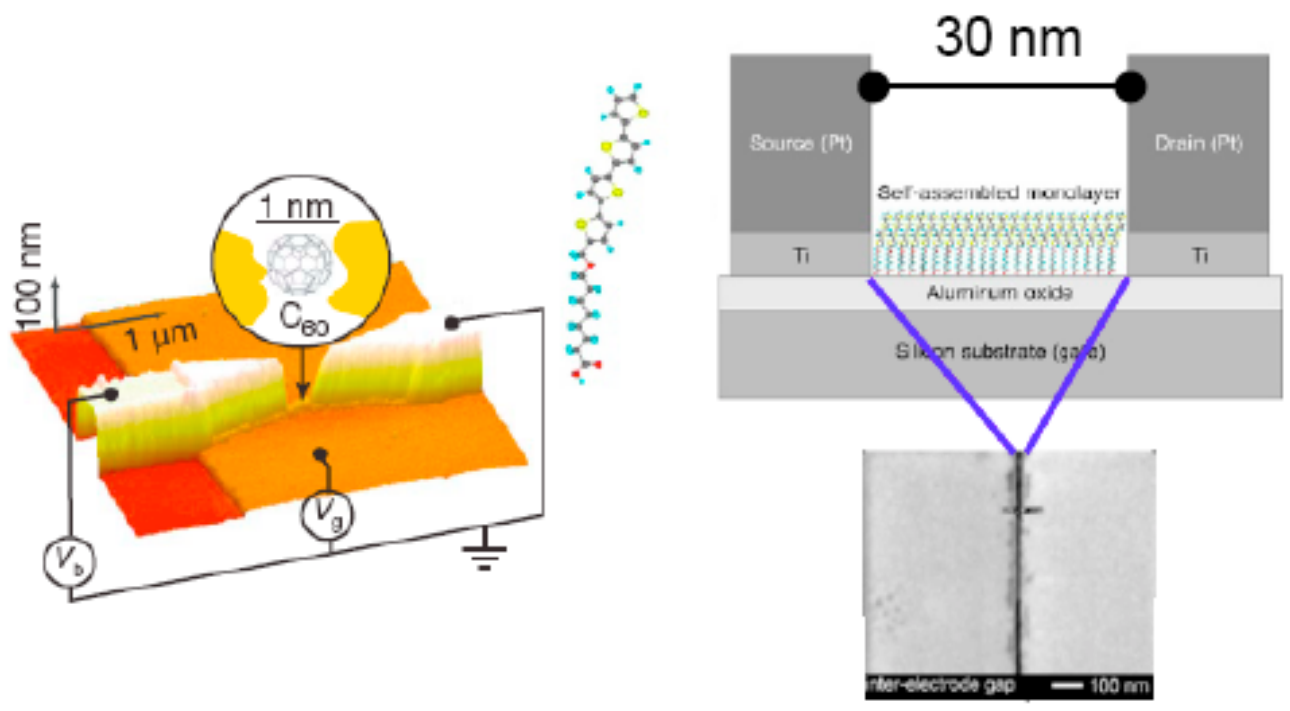

Fig. 8. Scheme of two molecular-based transistors, with a single molecule (left) [151] and with a self-assembled monolayers (right) [78].

At a single molecule level (single-molecule transistor), these devices have been used to study Coulomb blockade effects and Kondo effects at very low temperature. For instance, Coulomb blockade (electron flowing one-by-one between source and drain through the molecule due to electron-electron Coulomb repulsion, the molecule acting as a quantum dot) was observed for molecules such as fullerene ($C_{60}$) and oligo-phenyl-vinylene (OPV) weakly coupled to the source-drain electrodes.[152, 153] In this latter case, up to 8 successive charge states of the molecule have been observed [153]. With organo-metallic molecules bearing a transition metal, such as Cobalt terpiridynil complex and divanadium complex, Kondo resonance (formation of a bound state between a local spin on the molecule, or an island, or a quantum dot, and the electrons in the electrodes leading to an increase of the conductance at low bias, around zero volt) has also been observed in addition to Coulomb blockade.[83, 84] Kondo resonance is observed when increasing the coupling between the molecule and the electrodes (for instance by changing the length of the insulating tethers between the metal ion and the electrodes). Using such hybrid silicon-molecule transistor configuration, it was recently shown that it is possible to electrostatically modulate the current through the molecule by gating the molecular orbital with the underneath Si gate [154]. This hybrid silicon-molecule transistor configuration is also suitable to study and control the spin states and spin transport through molecules [151, 155-158].

At a monolayer level, self-assembled monolayer field-effect transistors (SAMFET) have been demonstrated at room temperature.[78, 159] The transistor effect is observed only if the source and drain length is lower than about 50 nm, that is, more or less matching the size of domains with well organized molecules in the monolayer. This is mandatory to enhance π stacking within the monolayer and to obtain a measurable drain current. SAM of tetracene, [159] terthiophene and quaterthiophene [78] derivatives have been formed in this nano-gap. Under this condition, a field effect mobility of about $3.5 \times 10^{-3}$ $cm^2V^{-1}s^{-1}$ was measured for a SAMFET made with a quaterthiophene (4T) moiety linked to a short alkyl chain (octanoic acid) grafted on a thin aluminum oxide dielectric. This value is on a par with those reported for organic transistor made of thicker films of evaporated 4T ($10^{-3}$ to $10^{-2}$ $cm^2V^{-1}s^{-1}$) [78]. The on/off ratio was about $2 \times 10^4$. For some devices, a clear saturation of the drain current vs. drain voltage curve has been observed, but usually, these output characteristics display a super linear behavior. This feature has been explained by a gate-induced lowering of the charge injection energy barrier at the source/organic channel interface.[76]. Recently, improvements in the fabrication and control of the structural organization of the molecules within the SAM have allowed extending this concept to 40 μm channel length SAMFET with improved mobility of $4 \times 10^{-2}$ $cm^2V^{-1}s^{-1}$ [160]. Such molecular devices are suitable for large area, flexible electronics, and a 15-bit code generator has been demonstrated with hundreds of SAMFETs addressed simultaneously.

## VI. Conclusion

This review describes several functions and devices that have been studied at the molecular scale. However, a better understanding and further improvements of their electronic properties are still mandatory and need to be confirmed. These results often suffer from lab-to-lab dispersion and more efforts are now required to improve reproducibility and repeatability. For viable applications, more efforts are also mandatory to test the integration of molecular devices with silicon-CMOS electronics (hybrid molecular-CMOS nanoelectronics). Moreover most of these devices are 2-terminals, a true/fully molecular 3-terminals device is still lacking. We have also pointed out that the molecule-electrode coupling and conformation strongly modify the molecular-scale device properties. Molecular engineering (changing ligand atoms for example) may be used to improve or adjust the electrode-molecule coupling. Albeit, many improvements have been recently obtained, a better control of the interface (energetics and atomic conformation) is still compulsory. Beyond the study of single or isolated devices, more works towards molecular architectures and circuits are required. Albeit not exclusive for molecular electronics, more new architectures must be explored (e.g. non von Neuman, neuronal and quantum computing…). For instance, the simplest instance of Shor's algorithm: factorization of $N = 15$, was implemented using seven spin-1/2 nuclei in a molecule as quantum bits [161], and other theoretical works investigate the possible use of molecules for quantum computing [162, 163]. Also, molecules, CNT and nanoparticles are suitable objects for the implementation of neuro-inspired devices [164, 165]. Open questions also concern the right approaches for inter-molecular device connections and nano-to-micro connections, the interface with the outer-world, hybridation with CMOS and 3D integration [166-170]. Beyond the CMOS probably asks to bet on devices non-based on the electron charge. Molecular devices using other state variables (e.g. spin, molecule configuration,…) to code a logic state are still challenging and exciting objectives.

## Acknowledgment

I thanks my colleagues, S. Lenfant, D. Guérin, N. Clément, S. Pleutin, from the "molecular nanostructures and devices" group at IEMN, and many others outside our group for fruitful collaborations and discussions.

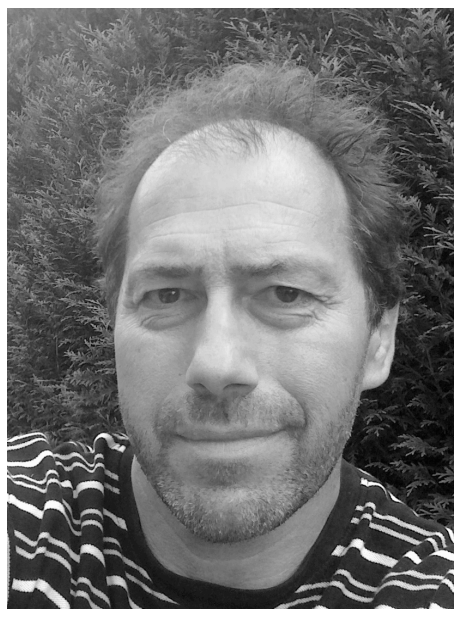

**Dominique Vuillaume** was born in 1956. He received the Electronics Engineer degree from the Institut Supérieur d'Electronique du Nord, Lille, France, 1981 and the PhD degree and Habilitation diploma in solid-state physics, from the University of Lille, France in 1984 and 1992, respectively. He is research director at the CNRS and he works at the Institute for Electronics, Microelectronics and Nanotechnology (IEMN), University of Lille. He is head of the « Molecular Nanostructures & Devices » research group at IEMN.

His research interests (1982-1992) covered Physics and characterization of point defects in semiconductors and MIS devices, Physics and reliability of thin insulating films, hot-carrier effects in MOSFET's. Since 1992, he has been engaged in the field of Molecular Electronics. His current research concerns:
- design and characterization of molecular and nanoscale electronic devices,
- elucidation of fundamental electronic properties of these molecular and nanoscale devices,
- study of functional molecular devices and integrated molecular systems,
- exploration of new computing paradigms using molecules and nanostructures.

He is the author or co-author of 150 scientific (peer-reviewed) papers in these fields. He was scientific advisor for industrial companies (Bull R&D center) and he is currently scientific advisor for the CEA "Chimtronique" research program.